\documentclass[aps,prd,twocolumn,preprintnumbers,floatfix,superscriptaddress]{revtex4-1}
\usepackage[final]{graphicx}
\usepackage{hyperref}
\usepackage{amsmath}
\usepackage{bbm}
\usepackage{amsfonts}
\usepackage{amssymb}
\usepackage{latexsym}
\usepackage{graphicx}
\usepackage[english]{babel}
\usepackage{multirow}
\usepackage{float}
\usepackage{url}
\usepackage{slashed}
\usepackage{xcolor}
\usepackage[utf8]{inputenc}
\usepackage{verbatim}
\usepackage{lipsum}

\newcommand{\be}{\begin{equation}}
\newcommand{\ee}{\end{equation}}
\newcommand{\ba}{\begin{array}}
\newcommand{\ea}{\end{array}}
\newcommand{\bea}{\begin{eqnarray}}
\newcommand{\eea}{\end{eqnarray}}

\newcommand{\nn}{\nonumber}

\makeatletter
\renewcommand{\maketag@@@}[1]{\hbox{\m@th\normalsize\normalfont#1}}%
\makeatother

\newlength{\halfpagewidth}
\divide\halfpagewidth by 2

\begin{document}
\title{Gravitational waves and monopoles dark matter from first-order phase transition}

\author{Jing Yang}
\affiliation{Department of Physics and Chongqing Key Laboratory for Strongly Coupled Physics, Chongqing University, Chongqing 401331, P. R. China}

\author{Ruiyu Zhou}\email{zhoury@cqupt.edu.cn}
\affiliation{School of Science, Chongqing University of Posts and Telecommunications, Chongqing 400065, P. R. China}

\author{Ligong Bian\footnote{Corresponding Author.}}
\email{lgbycl@cqu.edu.cn}
\affiliation{Department of Physics and Chongqing Key Laboratory for Strongly Coupled Physics, Chongqing University, Chongqing 401331, P. R. China}
\affiliation{ Center for High Energy Physics, Peking University, Beijing 100871, China}

\begin{abstract}

We study the possibility of monopoles serving as dark matter when they are produced during the
first-order phase transition in the dark sector. Our study shows that dark monopoles can contribute only a small piece of dark matter relic density within parameter spaces where strong gravitational waves can be probed by ET and CE, and the monopoles can contribute a sizable component of the observed dark matter relic density for fast phase transitions with short duration.

\end{abstract}

\maketitle

\section{Introduction}

Gravitational waves astronomy provides a new window to probe new physics beyond Standard Model that can accommodate first-order phase transitions~\cite{Caldwell:2022qsj}. The vector gauge bosons can serve as dark matter after the spontaneously broken of a hidden non-abelian gauge theory~\cite{Hambye:2008bq,Arina:2009uq,Carone:2013wla}.  Refs.~\cite{Ghosh:2020ipy,Prokopec:2018tnq,Baldes:2018emh} studied phase transition within hidden non-abelian gauge theories
and found the gravitational waves from the high-scale first-order phase transition can be probed by LIGO and future space-based gravitational wave detectors.
Ref.~\cite{Borah:2021ftr} investigated the possibility to search MeV-scale phase transition in such framework and found the parameter spaces that allowed first-order phase transition can be constrained by NanoGrav.  When the non-abelian gauge theory is broken to an abelian gauge theory with the vacuum manifold being $\pi_2(SU(2)/U(1))=\mathbb{Z}$,
dark matter might be 't Hooft-Polyakov
monopoles~\cite{tHooft:1974kcl,Polyakov:1974ek} that are produced during cosmological phase transitions~\cite{Kibble:1976sj,Zurek:1985qw}.
Monopoles serving as topological dark matter have been studied in Refs.~\cite{Nakagawa:2021nme,Graesser:2021vkr,Fan:2021ntg,Graesser:2020hiv,Daido:2019tbm,Sato:2018nqy,Kawasaki:2015lpf,Hiramatsu:2021kvu,GomezSanchez:2011orv,Nomura:2015xil,Nakagawa:2021nme,Murayama:2009nj,Evslin:2012fe,Terning:2019bhg,Khoze:2014woa,Baek:2013dwa,Bai:2020ttp}.

Previous studies~\cite{Khoze:2014woa,Baek:2013dwa,Bai:2020ttp} mostly focus on monopole dark matter from second-order phase transitions. We
focus on the dark monopoles formed when dark vacuum bubbles collided during the first-order phase transition~\cite{Einhorn:1980ik,Izawa:1982cu}.
We study heavy monopoles' contribution to dark matter relic density when the monopoles are produced during high-scale first-order phase transitions to be probed by LISA\,\cite{LISA:2017pwj}, TianQin\,\cite{TianQin:2015yph, Hu:2018yqb, TianQin:2020hid}, Taiji\,\cite{Hu:2017mde, Ruan:2018tsw}, DECIGO\,\cite{Seto:2001qf, Kudoh:2005as}, and BBO\,{\cite{Ungarelli:2005qb, Cutler:2005qq}, LIGO-Virgo~\cite{Thrane:2013oya,LIGOScientific:2014qfs,LIGOScientific:2016aoc,LIGOScientific:2019vic}, CE~\cite{Reitze:2019iox}, and ET~\cite{Punturo:2010zz,Hild:2010id,Sathyaprakash:2012jk}. We investigate the effects of phase transition duration and phase transition temperature on dark monopoles dark matter relic density.

\section{The dark $SU(2)_D$ phase transition}
The spontaneous breaking of a $SU(2)_D$ gauge theory with a scalar field in the adjoint representation may contain 't Hooft-Polyakov magnetic monopoles~\cite{tHooft:1974kcl} when the homotopy group satisfies
$\pi_2[SU(2)/U(1)] = \mathbb{Z}$. The relevant Lagrangian is given by,

\be\label{larg}
\mathcal{L}=-\frac{1}{4}F_{\mu \nu}^{a}F^{\mu \nu a}+\frac{1}{2}D_{\mu}\phi^a D^{\mu}\phi^a-V(\phi)\;.
\ee
Therein, the field strength of the $SU(2)$ field $A_{\mu}$ is
$
F_{\mu \nu}^{a}=\partial_{\mu}A_{\nu}^a-\partial_{\nu}A_{\mu}^a+g_D\epsilon^{abc}A_{\mu}^b A_{\nu}^c\;$, the kinetic term is
$
D_{\mu}\phi^a=\partial_{\mu}\phi^a+g_D\epsilon^{abc}A_{\mu}^b\phi^c$ with the $g_D$ being the gauge coupling constant and $\phi^a(a=1,2,3)$ being the adjoint scalars,  the tree-level potential at zero temperature is
\be
V(\phi)=-\frac{m^2}{2}\phi^a\phi^a+\frac{\lambda_{\phi}}{4}(\phi^a\phi^a)^2\;.
\ee
At zero temperature, the vacuum expectation value $v=w=m/\sqrt{\lambda_{\phi}}$ is the minimum of the potential $V$. After symmetry breaking, we get two massive gauge bosons $W_{\pm}'$ with mass $m_{W'}=g_D w$, one massless gauge boson $\gamma'$ and one massive scalar with $m_\phi=\sqrt{2}m$.

To investigate the phase transition process that leads to the symmetry breaking of $SU(2)_D\to U(1)_D$,
we
study the thermal effective potential at the one-loop level by considering high-temperature approximation~\cite{Quiros:1994dr}, which
  takes the form as~\cite{Khoze:2014woa}:
\be\label{veff}
V_T(\phi)=D(T^2-T_0^2)\phi^a \phi^a-ET(\phi^a \phi^a)^{3/2}+\frac{\lambda_T}{4}(\phi^a \phi^a)^2\;.
\ee
The relevant parameters are:
$
D=g_D^2/4, \ E=g_D^3/2\pi, \ T_0^2=\left(\sqrt{2}m^2-g_D^4w^2/2\pi\right)/(4D),
 \lambda_T=\lambda_{\phi}-3g_D^4/8\pi^2 \log(g_D^2w^2/a_BT^2)\;$ with $a_B\simeq e^{3.91}$.
Below the critical temperature, the phase transition would take place when at least one bubble is nucleated per horizon volume and per horizon time, which can be defined as\,\cite{Affleck:1980ac,Linde:1981zj,Linde:1980tt}:
\begin{eqnarray}\label{eq:bn}
\Gamma\approx A(T_n)e^{-S_3/T_n}\simeq 1\;.
\end{eqnarray}
Where $T_n$ is the nucleation temperature of the vacuum bubbles, and $S_3$ is the bounce action for an O(3) symmetric bounce solution that can be written as
\begin{eqnarray}
S_3(T)=\int 4\pi r^2d r\bigg[\frac{1}{2}\big(\frac{d \phi_b}{dr}\big)^2+V_T(\phi_b)\bigg]\;,
\end{eqnarray}
with $\phi_b = \phi$ in our case, and $V_T(\phi_b)$ is the thermal effective potential in Eq.\ref{veff}. The bubble nucleation events would be generated when one gets the bounce solution from solving the equations of motion for $\phi_b$:
\begin{eqnarray}
\frac{d^2\phi_b}{dr^2}+\frac{2}{r}\frac{d\phi_b}{dr}-\frac{\partial V(\phi_b)}{\partial \phi_b}=0\;,
\end{eqnarray}
with the boundary conditions being
\begin{eqnarray}
\lim_{r\rightarrow \infty}\phi_b =0 \;, \quad \quad {\left. {\frac{{d{\phi _b}}}{{dr}}} \right|_{r = 0}} = 0\;.
\end{eqnarray}
One typical parameter is the phase transition strength $\alpha$, which can be calculated  based on the trace of the energy-momentum tensor~\cite{Giese:2020rtr,Giese:2020znk,Guo:2021qcq}:
\begin{equation}
\alpha = \frac{1}{3 \omega_s}((1+\frac{1}{c^{2}_s}) \Delta V_{eff} - T\frac{d \Delta V_{eff}}{dT})|_{T=T_n}\;,
\end{equation}
where $\omega$ is the enthalpy density, and subscripts $``s"$ indicate the quantities outside the bubbles. $\Delta V_{eff}$ is the thermal effective potential difference between the symmetric and broken phase.
The speed of sound $c_{s}$ is defined as:
$
c_{s}^2 = (dp/dT)/(de/dT)
$,
where $e$ ($p$) is the energy density (the pressure).
Another typical parameter $\beta$ which characterizes the inverse duration of the first-order phase transition (in units of Hubble) can be obtained as
$
\beta/H_n=\left.T d (S_3(T)/T)/(d T)\right\vert_{T=T_n}\;
$
with $H_n$ being the Hubble constant at the nucleation temperature $T_n$.

 \begin{figure}[!htp]
\begin{center}
\includegraphics[width=0.22\textwidth]{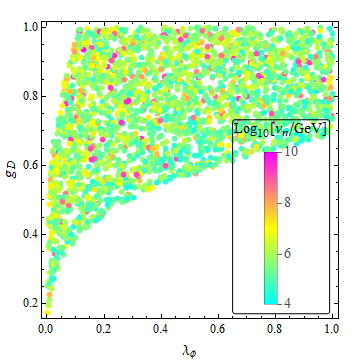}
\includegraphics[width=0.22\textwidth]{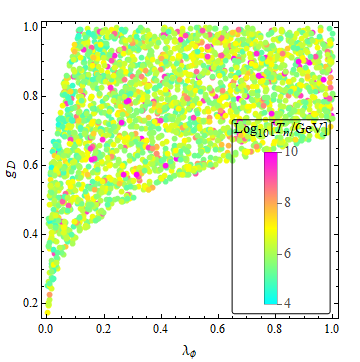}
\includegraphics[width=0.22\textwidth]{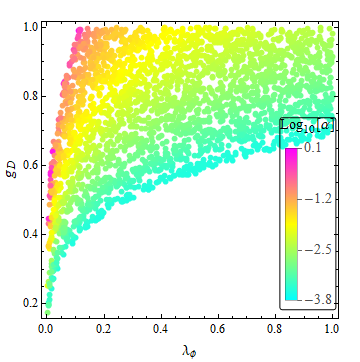}
\includegraphics[width=0.22\textwidth]{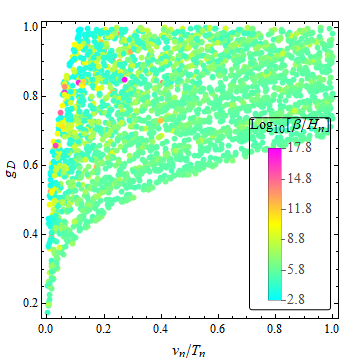}
\caption{The top two plots show the $v_n$ and $T_n$ for phase transitions with different model parameters; and the bottom two plots depict the values of phase transition strength $\alpha$ and inverse phase transition duration $\beta/H_n$.  } \label{ptpatterns}
\end{center}
\end{figure}

With the above procedure, we scan model parameters of $g_D \sim [0,1],
w\sim [10^5,10^{10}]$ GeV, $\lambda_\phi\sim[0,1]$ to find bounce solutions by utilizing {\it FindBounce}~\cite{Guada:2020xnz}, and calculate the phase transition strength and duration. Our results are shown in Fig.~\ref{ptpatterns}.
The top two plots present the distributions of the background field value ($v_n$) at the nucleation temperature ($T_n$), both $v_n$ and $T_n$ range from $10^4$ GeV to $10^{10}$ GeV. The yielded phase transition strength $\alpha$ and the inverse phase transition duration ($\beta/H_n$) are shown in the bottom two panels respectively.
In most parameter spaces, we have weak phase transition with $\alpha<0.1$ and large $\beta/H_n$, where stronger phase transition strength $\alpha$ can be obtained with smaller $\lambda_\phi$ and larger $g_D$. As will be studied later, dark monopoles generated during the phase transition with larger $\beta$ can contribute more dark matter relic density, and the strong gravitational wave requires relatively small $\beta/H_n$ and strong phase transition with a large $\alpha$.

 \begin{figure}[!htp]
\begin{center}
\includegraphics[width=0.3\textwidth]{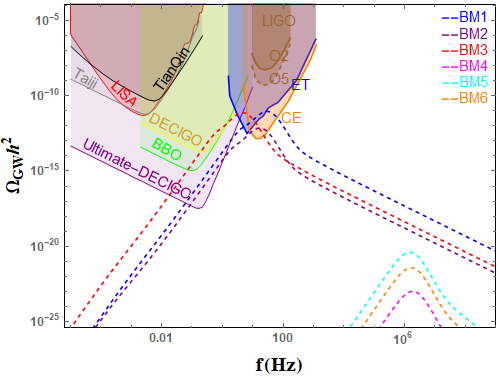}
\caption{Gravitational wave spectra for phase transition benchmarks in Table.~\ref{ptpar}.  } \label{gwmon}
\end{center}
\end{figure}

A first-order PT occurs when dark vacuum bubbles nucleated and merged with each other as the temperature of the Universe drops. Gravitational waves  could then be generated, with the spectrum of gravitational wave can be obtained systematically~\cite{Caprini:2019egz}. Generically, the prediction of the gravitational wave spectrum depends on four crucial parameters: the bubble wall velocity $v_w$, the phase transition temperature (we take $T_n$ in this study), the phase transition strength $\alpha$, and the inverse phase transition duration $\beta$.
In this work, we consider three sources of gravitational waves during the dark first-order PT:  1) the bubble collision, here we take the widely used envelop approximation ~\cite{Kosowsky:1991ua,Kosowsky:1992rz,Kosowsky:1992vn,Kamionkowski:1993fg,Huber:2008hg} (see also~\cite{Child:2012qg} and analytic estimations~\cite{Jinno:2016vai})\footnote{However, recent numerical simulations found that the scalar oscillation stage would continue contributing to GW radiation, see Refs.~\cite{Cutting:2018tjt,Cutting:2020nla,Di:2020ivg,Zhao:2022cnn}.}; 2) the sound waves in the plasma~\cite{Hindmarsh:2013xza,Hindmarsh:2015qta}, and 3) the magnetohydrodynamic turbulence (MHD)~\cite{Hindmarsh:2013xza,Hindmarsh:2015qta}, which may subdominant when a small fraction of the energy flows into the MHD~\cite{Caprini:2009yp,Binetruy:2012ze}.
In Fig.~\ref{gwmon}, we present the gravitational wave predictions for several benchmarks given in Table.~\ref{ptpar}. Where we find that large $\beta/H_n$ (and small $\beta/H_n$) yield low (and high) magnitude of GWs. The possibility of the monopoles serving as DM in these benchmarks will be studied below.

\begin{table}[!htp]
\begin{center}
\begin{tabular}{c  c  c  c  c  c  c  c  c}
\hline
BM& ~~$g_D$~~&$~~\lambda_{\phi}$~~&~~$w(\text{GeV})$~~&$~~T_{n}(\text{GeV})$~~&~~$\alpha$~~~~&~~$\beta/H_{n}$~~\\
 \hline
 $BM_1$ &  $0.99$    &  $0.11$   &    $2.77\times 10^6$   &   $3.21\times 10^5$ & $1.28$ & $754.07$ \\
 $BM_2$ &  $0.64$    &  $0.02$ &    $6.82\times 10^5$   &   $6.31\times 10^4$ & $0.90$ & $2213.46$ \\
 $BM_3$ &  $0.57$    &  $0.01$   &    $6.77\times 10^5$   &   $5.97\times 10^4$ & $0.81$ & $640.42$  \\
 $BM_4$ &  $0.66$    &  $0.63$   &    $4.87\times 10^7$   &   $6.97\times 10^7$ & $4.26\times10^{-4}$ & $2.20\times10^5$ \\
 $BM_5$ &  $0.92$    &  $0.61$   &    $6.93\times 10^7$   &   $6.88\times 10^7$ & $5.12\times10^{-3}$ & $1.79\times10^5$ \\
 $BM_6$ &  $0.87$    &  $0.70$   &    $4.16\times 10^7$   &   $4.74\times 10^7$ & $2.73\times10^{-3}$ & $2.87\times10^5$ \\
 \hline
\end{tabular}
\caption{The benchmark points for gravitational waves and dark monopoles.}\label{ptpar}
\end{center}
\end{table}

\section{Monopole dark matter}
When the phase transition occurs, monopoles can be formed with a rate of $p\sim \mathcal{O}(10^{-1})$ after dark vacuum bubbles collided with each other.  The monopole density $n_{m,*}$ can be related to the bubble number density $n_{b,*}$ through the following relation~\cite{Einhorn:1980ik},
\be
n_{m,*}=p n_{b,*}=p\frac{\beta^3}{8\pi v_w^3}\;.
\ee
When monopoles are formed, configurations of the dark scalar and dark gauge fields are as follows~\cite{Hiramatsu:2021kvu}:
\bea
\phi^a&=&v_nH(r)\frac{x^a}{r}\;,\nn\\
A_{i}^{a}&=&\frac{1}{g_D}\frac{\epsilon^{aij} x^j}{r^2}F(r)\;,   \  \  (i,j=1,2,3),
\eea
where $\epsilon^{aij}$ is the anti-symmetric tensor with a convention $\epsilon^{123}=1$, and $r=\sqrt{x^2+y^2+z^2}$.
In terms of the above configurations, the Lagrangian given in Eq.\ref{larg} reduces to
\bea
L&=& \int \mathcal{L}d^3x=-4\pi \int_{0}^{\infty}r^2 dr \bigg[\frac{1}{g_D^2 r^2}\left(\frac{dF}{dr}\right)^2  \nn \\
&+& \frac{2F^2(1-F)}{g_D^2 r^4}+\frac{F^4}{2g_D^2 r^4}+\frac{v_n^2}{2} \left(\frac{dH}{dr} \right)^2   \nn \\
 &+& \frac{v_n^2 H^2}{r^2} (1-F)^2+V_T(H)\bigg]\;.
\eea
Introducing the dimensionless variable $\xi=v_n r$, we have
\bea
L&=&-\frac{4\pi v_n}{g_D^2} \int_{0}^{\infty} d\xi \bigg[\left(\frac{dF}{d\xi}\right)^2+\frac{2F^2(1-F)}{ \xi^2} \nn \\
&+&\frac{F^4}{2 \xi^2}+\frac{g_D^2 \xi^2}{2} \left(\frac{dH}{d\xi} \right)^2+g_D^2H^2(1-F)^2 \nn \\
&+&\frac{g_D^2 V_T(H)}{v_n^4}\xi^2 \bigg].
\eea
Then, equations of motion of the system can be obtained as:
\bea\label{eom}
&&\frac{d^2F}{d\xi^2}=\frac{F}{\xi^2}(1-F)(2-F)+g_D^2H^2(F-1)\;,\\
&&\frac{d^2H}{d\xi^2}+\frac{2}{\xi}\frac{dH}{dx}=\frac{2H}{\xi^2}(1-F)^2+\frac{1}{v_n^4}\frac{dV_T(H)}{dH}\;.
\eea
Where, the functions $H(\xi)$ and $F(\xi)$ satisfy boundary conditions:
\bea
&&\lim_{\xi\to 0}H(\xi) \rightarrow 0\;, \ \ \ \ \lim_{\xi \to \infty} H(\xi) \rightarrow 1\;,\\
&&\lim_{\xi\to 0 }F(\xi) \rightarrow 0\;, \ \ \ \   \lim_{\xi\to \infty }F(\xi) \rightarrow 1\;.
\eea

After the $H(\xi)$ and $F(\xi)$ are solved numerically, we can obtain the mass of the monopole as
 \bea\label{mm}
 M_m&=&\int [-\mathcal{L}-V_T(v_n)]d^3x \nn \\
 &=&\frac{4\pi v_n}{g_D^2} \int_{0}^{\infty} d\xi \bigg[\left(\frac{dF}{d\xi}\right)^2+\frac{2F^2(1-F)}{ \xi^2} \nn \\
&+&\frac{F^4}{2 \xi^2}+\frac{g_D^2 \xi^2}{2} \left(\frac{dH}{d\xi} \right)^2+g_D^2 H^2(1-F)^2 \nn \\
&+&\frac{g_D^2 V_T(H)}{v_n^4}\xi^2-\frac{g_D^2 V_T(H=1)}{v_n^4}\xi^2 \bigg].
\eea

The relic density of dark monopoles can be calculated as
\be
\Omega_{M}=\frac{\rho_{M,0}}{\rho_{crit,0}}=\frac{M_m n_{m,*}s_0}{3M_{pl}^2H_0^2 s_{*}},
\ee
where $s_{*}$ is the entropy density of the universe at the temperature of phase transition, $s_0$ is the entropy density today and $H_0$ is the current Hubble constant.
Utilizing the relation $s(T)=\frac{2\pi^2}{45}g_s(T)T^3$, we have the dark monopole dark matter relic density
\be
\Omega_{DM}h^2=pM_m\frac{\beta^3g_s(T_0)T_0^3h^2}{24\pi v_w^3 g_s(T_n)T_n^3M_{pl}^2H_0^2},
\ee
where $g_s$ is the effective number of degrees of freedom in entropy.

\begin{figure}[!htp]
\begin{center}
\includegraphics[width=0.22\textwidth]{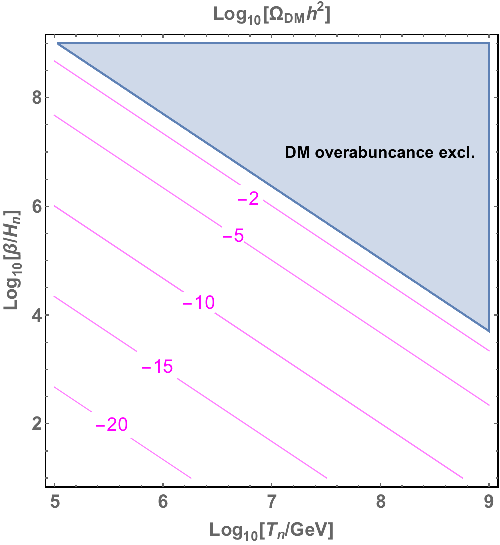}
\includegraphics[width=0.24\textwidth]{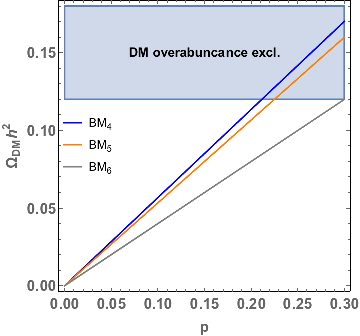}
\caption{Monopoles dark matter relic density. In left panel, we present the monopoles dark matter relic abundance as functions of $\beta$ and phase transition temperature $T_n$ with bubble velocity $v_w=1$
and monopole generation rate $p=0.1$. In the right plot, we show the dark matter relic density for three benchmarks in Table.~\ref{ptpar}.   } \label{dmmon}
\end{center}
\end{figure}

In the left plot of Fig.~\ref{dmmon}, for illustration, we take $M_m\sim 4 \pi v_n/g_D^2\sim 4 \pi T_n$ since most phase transition points concentrate around $v_n/T_n\sim 1$ with $g_D\sim [0.5,1]$. The figure depicts that the dark matter relic density increases with increase of $\beta/H_n$ and the phase transition temperature $T_n$. The parameter spaces with long phase transition duration (i.e., a small $\beta/H_n$) that can produce strong gravitational waves (see $BM_{1,2,3}$ in the Fig.~\ref{gwmon}) yield negligible contributions to the dark matter relic density.
The right plot shows that the short phase transition duration cases of $BM_{4,5,6}$ (with large $\beta/H_n$) can contribute even most part of the observed dark matter relic density depending on the monopole generation rate when dark vacuum bubbles collided with each other. Where,
the monopole mass was calculated through Eq.~\ref{mm} after we obtained
the profiles of $H(\xi)$ (solid) and $F(\xi)$ (dotted) for the $BM_{4,5,6}$ that satisfy Eq.~\ref{eom}, see Fig.~\ref{prof}.
Fig.~\ref{prof} shows that both $H(\xi)$ and $F(\xi)$ approach unity for large $\xi$.
After the $SU(2)_D$ is broken to a $U(1)_D$ theory, the dark photon $\gamma^\prime$ didn't get mass during the phase transition, which will contribute to the effective neutrino number.
Following Refs.~\cite{Ghosh:2020ipy,Khoze:2014woa}, we obtain
$
\Delta N_{\text{eff}}(T_{\text{CMB}})\approx 0.055,
\Delta N_{\text{eff}}(T_{\text{BBN}})\approx 0.202
$ that will be tested by future CMB Stage IV experiments~\cite{Borah:2021ftr}.

\begin{figure}[!htp]
\includegraphics[scale=0.5]{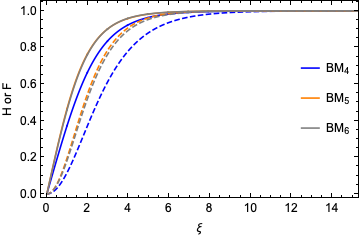}
\caption{ The profiles of $H(\xi)$(solid lines) and  $F(\xi)$(dashed lines). The blue, orange, and gray lines correspond to three benchmarks in the Table.~\ref{ptpar} respectively.  }\label{prof}
\end{figure}

\section{Conclusion and discussions}

In this paper, we investigated the possibility of dark monopoles serving as dark matter and its associated gravitational waves production during the dark first-order phase transition.
Our study shows that the monopoles produced during the phase transition can contribute part of the dark matter relic density depending on phase transition properties.
We find heavy monopoles produced from short duration high-scale phase transitions, that produce weak gravitational waves, can contribute considerable components of the dark matter relic density. Meanwhile, we observe that for the parameter spaces with strong gravitational waves to be probed by CE and ET, the monopoles can only contribute a negligible piece of the observed dark matter relic density.

In this paper, we didn't consider large production rates of monopoles therein one needs to consider monopole-anti-monopole annihilation for the study of dark matter relic density, see Refs.~\cite{Khoze:2014woa,Baek:2013dwa,Bai:2020ttp} for example. The exact production of the monopoles would rely on the bubbles merging process during the first-order phase transition, we left the study to future lattice simulations.

\section{Acknowledgement}

The work is supported by the National Key Research and Development Program of China Grant No. 2021YFC2203004. The work of Ligong Bian is supported in part by the National Natural Science Foundation of China under the grants Nos.12075041, 12047564, the Fundamental Research Funds for the Central Universities of China (No. 2021CDJQY-011 and No. 2020CDJQY-Z003), and Chongqing Natural Science Foundation (Grants No.cstc2020jcyj-msxmX0814).

\bibliography{mn}

\end{document}